\documentclass[prl,aps,showpacs,groupedaddress,twocolumn,floatfix]{revtex4}

\usepackage{amsmath,amsfonts,amssymb}
\usepackage{wrapfig}
\usepackage{graphicx}
\usepackage{bbm}
\usepackage{graphics}

\def \beq {\begin{equation}}
\def \eeq {\end{equation}}
\def \ba {\begin{eqnarray}}
\def \ea {\end{eqnarray}}
\newcommand{\upp}{\hspace{-0.2 pt}\uparrow}
\newcommand{\downn}{\hspace{-0.2 pt}\downarrow}

\newcommand{\ketbrad}[1]{|#1\rangle\!\langle #1|}

\newcommand{\mean}[1]{\langle#1\rangle}

\def\ket#1{\left| #1\right>}
\def\bra#1{\left< #1\right|}
\renewcommand{\section}[1]{}

\begin{document}

\title{Cavity quantum electrodynamics with semiconductor double-dot
molecules on a chip}
\author{J. M. Taylor and  M. D. Lukin}
\affiliation{Harvard University, Department of Physics, 17 Oxford
  Street, Cambridge, MA  02138 USA}
\begin{abstract}

  We describe a coherent control technique for coupling electron spin
  states associated with semiconductor double-dot molecule to a
  microwave stripline resonator on a chip. We identify a novel regime
  of operation in which strong interaction between a molecule and a
  resonator can be achieved with minimal decoherence, reaching the
  so-called strong coupling regime of cavity QED. We describe
  potential applications of such a system, including low-noise
  coherent electrical control, fast QND measurements of spin states,
  and long-range spin coupling.

\end{abstract}


\maketitle

Controlling quantum behavior of realistic solid-state systems is an
intriguing challenge in modern science and engineering. While over the
past several decades much progress has been made in manipulation of
isolated atomic and optical systems, only recently have solid-state
systems demonstrated controlled, coherent behavior at a single quantum
level~\cite{nakamura99,mooij99,vion02,martinis02,wallraff04,petta05}.
The complex environment of solid state systems make them significantly
more challenging to isolate and operate coherently.  Furthermore,
robust quantum control techniques analogous to those used in AMO
physics still need to be developed.

Recently, a novel approach to quantum manipulation of spin-based
quantum systems in semiconductor quantum dots has been proposed and
experimentally realized~\cite{levy02,petta05,taylor06}. This
approach combines spin and charge manipulation to take advantage of
the long memory times associated with spin states and, at the same
time, to enable efficient readout and coherent manipulation of coupled
spin states using intrinsic interactions. While this allows one to
consider architectures based on pulsed quasi-static electrical control
and static magnetic field~\cite{taylor05nature}, for many purposes it
is desirable to develop fast and robust quantum control techniques
based on microwave manipulation.

In this Letter we describe a technique for electrical coupling of
electron spin states associated with semiconductor double-dot molecule
to microwave stripline resonator on a
chip~\cite{childress04,blais04,wallraff04}. The essential idea is to
use an effective electric dipole moment associated with exchange
coupled spin states of a double-dot molecule to couple to the
oscillating voltage associated with a stripline resonator. Taking into
account the main decoherence mechanisms for both spin and charge
degrees of freedom, we identify an optimal point of operation in which
strong interaction between molecule and resonator can be achieved with
minimal decoherence, thereby enabling the strong coupling regime of
cavity QED. Finally, we describe potential applications including
low-noise coherent electrical control, fast QND measurements of spin
states, and long-range coupling of pairs of spins.

Before proceeding we note the early proposals for achieving the strong
coupling regime of cavity QED with Cooper pair qubits and single
electrons in double-dot molecules~\cite{childress04,blais04}. Strong
coupling of superconducting qubits to stripline cavities have been
recently implemented in pioneering experiments by Schoelkopf and
co-workers~\cite{wallraff04}. This system has enabled a range of
beautiful demonstrations ranging from control and measurement of the
qubit through the resonator to SWAP of the qubit with the state of the
resonator.  Finally, ideas similar to the present work have been
proposed recently, using vertical quantum dot systems~\cite{burkard06}.

\begin{figure}
\includegraphics[width=3.0in]{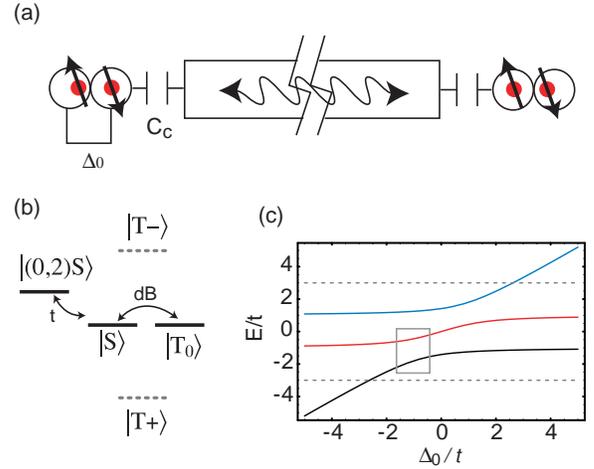}
\caption{
(a) Schematic of two double dots, biased with external potential
$\Delta_0$, capacitively coupled to a transmission line resonator.
(b) Energy level diagram showing the (0,2) singlet and the four (1,1)
two-spin states.
(c) Low energy spectrum in units of $t$ with small gradient $dB=t/10$: for large, negative
$\Delta_0$, the ground state is $\ket{(0,2)S}$.  The $T_0$ triplet
(red) and $T_{\pm}$ triplets (dashed grey) are far from resonance with the $(0,2)$
triplet.  Note the optimal point  (gray box) occurs at the left-most avoided crossing.
\label{f:picture}}
\end{figure}

\section{Effective hamiltonian}
We consider the specific system outlined in the
Figure~\ref{f:picture}. Here two electron spins are localized in
adjacent quantum dots
, coupled via tunneling.  One of the dots is capacitively coupled to a
transmission line resonator.  A modest (100 mT) external magnetic
field along an axis $z$ Zeeman-splits the spin-aligned states,
$\ket{T_+}=\ket{\upp \upp}$ and $\ket{T_-}=\ket{\downn \downn}$, while
the spin-anti-aligned states are used as a qubit degree of freedom:
$\ket{(1,1)T_0} = (\ket{\upp \downn} + \ket{ \downn \upp})/\sqrt{2}$
and $\ket{(1,1)S} = (\ket{\downn \upp}-\ket{ \downn \upp})/\sqrt{2}$.
The notation $(n_L,n_R)$ labels the number of electrons in the left
and right quantum dots.  In addition to the qubit states, an auxiliary
singlet state with two electrons in one quantum dot, $\ket{(0,2)S}$,
is coupled via tunneling $t$ to the separated singlet, $\ket{(1,1)S}$
(Fig.~\ref{f:picture}b).  The energy of the auxiliary state is
determine largely by the bias $\Delta$ due to the electric field;
control of $\Delta$ allows to control the qubit's evolution.  The
Hamiltonian for this three state system is:
\begin{eqnarray}
  H_{DD} & = & \Delta \ketbrad{(0,2)S} + t \ket{(1,1)S}\bra{(0,2)S} + 
\nonumber \\ 
& & dB \ket{(1,1)S}\bra{(1,1)T_0} +
  {\rm H.c.} . \label{e:hdd}
\end{eqnarray}
We have introduced a static magnetic field gradient between the two
dots $dB = g^* \mu_B (B_z^L - B_z^R)$, which mixes the singlet and
triplet states.  This system has been studied in detail in
Refs.~\onlinecite{brandes03,childress04,coish05,taylor06}.  Parameter
ranges for experiments are $t \simeq 0-10 \mu$eV and $dB \sim 0.1-1
\mu$eV.  State preparation, measurement, 1-qubit gates, and local
2-qubit gates can be achieved by changing the bias $\Delta$ between
$\ket{(0,2)S}$ and
$\ket{(1,1)S}$~\cite{levy02,taylor05nature,petta05}.

Interaction with the resonator is included naturally by writing the bias as a
contribution from static fields, $\Delta_0$, set by voltages on the
gates defining the double dot, and a contribution from the resonator
itself:
\begin{equation}
  \Delta = \Delta_0 + e \hat{V} \frac{C_c}{C_{\rm tot}}\ ,
\end{equation}
where $C_c$ is the capacitive coupling of the resonator to the
dot, while $C_{\rm tot}$ is the total capacitance of the double dot.
The voltage due to the resonator of length $l$, capacitance per
unit length $C_0$, and impedance $Z_0$ is quantized as~\cite{childress04}
\begin{equation}
  \hat{V} = \sum_k \sqrt{\frac{\hbar \omega_k}{l C_0}}(\hat{a}_k +
  \hat{a}_k^{\dag}) \ .
\end{equation}

We now deduce an effective Hamiltonian for the system when the
splitting between eigenstates of Eq.~\ref{e:hdd}, $\omega$, is
comparable to the fundamental mode frequency of the resonator,
$\omega_0 = \pi / l Z_0 C_0$ (i.e., the cavity detuning $\delta =
\omega-\omega_0$ is small).  Neglecting the higher energy modes, we
write the Hamiltonian for the resonator itself as $H_{TLR} = \hbar
\omega_0 \hat{a}^{\dag} \hat{a}$, where $\hat{a}$ is the lowest energy
mode destruction operator, and the interaction with the dot is
\begin{equation}
  U = g (\hat{a} + \hat{a}^{\dag}) \ketbrad{(0,2)S}
\end{equation}
where $g = e \frac{C_c}{C_{\rm tot} l C_0} \sqrt{\hbar \pi/ Z_0} =
\eta \omega_0$ is the vacuum Rabi coupling between the double dot and
the resonator.  In practice, for $\omega_0 = 2 \pi \times 10$ GHz, $g
= 2 \pi \times
100$ MHz (i.e., $\eta = 10^{-2}$) is
achievable~\cite{childress04,wallraff04,blais04}).  A reduction of
$\omega_0$ by lowering $l C_0$ results in a comparable reduction of
$g$.  Finally, as $Q > 10^6$ resonators are possible in the microwave
domain~\cite{childress04,wallraff04}, the
quantization of the resonator voltage is appropriate.  In practice,
the need to work with a finite local external magnetic field (100 mT)
may limit $Q \approx 10^4$~\cite{frunzio05}.

\section{Operating point}
We seek a set of parameters $\{\Delta_0,T_0,dB\}$ such that the system
can be coupled to and controlled by the resonator.  This ``operating
point'' should maximize the coupling to the resonator and minimize the
noise in the combined system.  To motivate the optimal choice, let us
first proceed with small tunnel coupling $t$ and cavity coupling $g$.
In this case the eigenstates are $\ket{\downn \upp}, \ket{(0,2)S},
\ket{\upp \downn}$, with eigenenergies $-dB, \Delta_0, dB$.  Setting
the energy difference $\tilde{\Delta} = dB - \Delta_0 \approx 0$,
degenerate perturbation theory in the tunnel coupling $t$ reveals an
avoided crossing at this balanced point between $\ket{\downn \upp}$
and $\ket{(0,2)S}$ with an energy gap $\omega = \sqrt{\tilde{\Delta}^2
  + 4 t^2}$ around $\tilde{\Delta}=0$ (Fig.~\ref{f:picture}b) and
mixing angle $\theta = \frac{1}{2}\tan^{-1}(2 t/\tilde{\Delta})$.
Working in the rotating frame with the rotating wave approximation, we
find an effective Hamiltonian for the combined system to be
\begin{equation}
  H = \hbar (\omega - \omega_0) \ketbrad{1} + g_{\rm eff} \hat a \ket{1}\bra{0} +
  {\rm H.c.}
\end{equation}
where, at $\theta = \pi/4$, $\ket{0(1)} = (\ket{\downn \upp} -(+)
\ket{(0,2)S})/\sqrt{2}$ and
\begin{equation}
  g_{\rm eff} = -\frac{1}{2} \eta \omega_0 \sin 2\theta \approx - \eta t\ .
\end{equation}

We choose $\theta=\pi/4$ for two reasons.  First, at this point the
gap $\omega$ is insensitive to first-order changes in $\Delta_0$ and
$dB$. Such zero-derivative points are optimal for controlling quantum
bits with imperfect electronics and in the presence of low frequency
noise~\cite{makhlin99,vion02,simmonds04,%
  falci05,yao05}.  In our case this suppresses fluctuations in control
electronics ($\Delta_0$) and in the magnetic field gradient (due to
hyperfine interactions), which are the dominant dephasing terms.  Second,
the coupling coefficient $g_{\rm eff}$ is directly proportional to the
energy splitting in the system, as would be expected of a ``bare'' coupling
of comparable strength.

The cost of working at this point is increased sensitivity to
fluctuations in $t$ (it effects both $g_{\rm eff}$ and $\omega$ at
first order) and a higher probability of relaxation due to the
enhanced charge admixture of the qubit states (see the analysis of
inelastic effects below).  The former could be mitigated: as
no time-dependent control of $t$ is required, the gates that set $t$
can be heavily filtered to greatly reduce potential noise.

For long term storage of the quantum information, we can adiabatically
map the states $\ket{0},\ket{1}$ to the spin states $\ket{\downn
  \upp}, \ket{\upp \downn}$ via change of $\Delta_0$, going to a well
separated regime.  As long as the change of $\Delta_0$ is slow with
respect to the splitting $dB$, the process is entirely adiabatic;
furthermore, we are only sensitive to charge fluctuations on a
frequency scale faster than $\sim dB$.  The spin states, protected by
occassional refocusing, are mostly insensitive to charge relaxation
and dephasing~\cite{loss98,burkard99}.  Thus, when long quantum memory
times or insensitivity to charge-based relaxation is necessary, the
system can be adiabatically mapped to such a separated regime; when
operation with the resonator is necessary, they can be mapped back to
the balanced regime of $\Delta_0 \approx - dB$.

\section{Generalized approach}
We now proceed to generalize the above analysis to an arbitrary set of
parameters $p = \{dB,t,\Delta_0\}$, treating the interaction with the
cavity as a perturbation. We write the eigenenergies of $H_{DD}$ as
$E_m(p)$ ($m=0,1,2$, and $E_0 \le E_1 \le E_2$) and the eigenvectors
as
\begin{equation}
 \left( \begin{array}{c}
\ket{0} \\
\ket{1} \\
\ket{2} 
\end{array}\right)
=
\chi^{\dag}(p)
 \left( \begin{array}{c}
\ket{(1,1)T_0} \\
\ket{(1,1)S} \\
\ket{(0,2)S} 
\end{array}\right)
\end{equation}
where $\chi$ is a 3x3 unitary matrix.  
In the rotating wave approximation, we may neglect all cavity
interaction terms diagonal in the eigenstates.  Furthermore, when the
energy differences are non-degenerate ($E_1-E_0 \neq E_2 - E_1$), we
may also only keep coupling to the transition with energy $E_1 - E_0
\approx \hbar \omega_0$.  Then,
\begin{equation}
  U = g \hat a^{\dag} \chi_{02} \chi_{21}^* \ket{0}
  \bra{1} + {\rm H.c.} 
\end{equation}
Thus, $g_{\rm eff}(p) = g \chi_{02}(p) \chi_{21}^*(p)$.  Keeping $t$
fixed, we plot $g_{\rm eff}$ and the expected error in operations due
to relaxation (calculated below).  Of immediate interest is the regime
of $dB \sim t$, where $g_{\rm eff}/g > 0.3$ and $\xi \neq 0$.  This
suggests that maximal coupling to the cavity occurs when $t,
-\Delta_0$, and $dB$ are all comparable.

\begin{figure}
\includegraphics[width=3.0in]{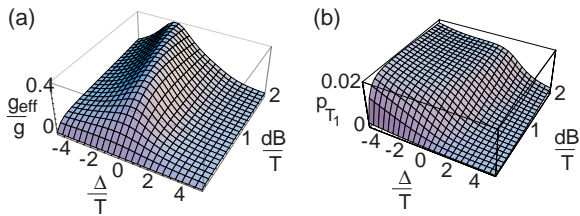}
\caption{
(a) Coupling strength, $g_{\rm eff}$, in units of bare coupling $g$,
as a function of field gradient $dB$ and bias $\Delta_0$.  (b)
Calculated probability of error in swapping a double-dot state with
the resonator state due to charge-based relaxation using $t=4\ \mu$eV and an inverse
temperature of 10 $\mu$eV.
\label{f:coupling}}
\end{figure}

\section{Noise and decoherence}
We now demonstrate the favorable noise properties of the balanced system.
Three kinds of error are considered: relaxation of the charge system
in a time $T_1$, additional dephasing of the charge system in a time
$T_{2,a} = T_2 - 2 T_1$, and decay of the cavity photon at a rate
$\kappa$.  We assume the additional dephasing arises from low
frequency
fluctuations~\cite{vion02,martinis03,simmonds04,yao05,taylor05nature,taylor06},
due to changes of the electrostatic gates and magnetic field gradient,
e.g., as noise on $\Delta_0$ and $dB$.

For charge-based relaxation, coupling to a phonon bath with spectral
density $\rho(\omega) = \sum_{k} |g_k|^2 \delta(\omega-\omega_k)$ will
lead to decay~\cite{caldeira83}.  Using the spin-boson model in the
perturbative regime yields an overall error rate from relaxation and
incoherent excitation:
\begin{equation}
  1/T_1 = 2 \pi \rho(\omega_{10}) |\chi_{20}\chi_{12}|^2
  \coth(\omega_{10} \beta/2)\ .
\end{equation}
We estimate $\rho(\omega)$ at $\omega = 76 \mu$eV/$\hbar$ using
measured charge relaxation times~\cite{hayashi03,petta04}.  For a
double quantum dot system in GaAs and small $\omega$, $\rho(\omega)
\propto \omega^3$~\cite{brandes03}.  This indicates $2 \pi
\rho(\omega) \approx \hbar^2 \omega^3 / (1 {\rm meV})^2$ for the low
energy limit.  Reducing the resonant frequency lowers relaxation
rates.  Near the optimal point, with $\omega_{10} \sim 2\pi \times$ 2
GHz, $T_1 \sim 1 \mu$s.

Additional dephasing ($T_{2,a}$ term) arises from variations of the
energy gap, $\Delta_0(t) = \Delta_0 + \epsilon(t)$ with
$\mean{\epsilon(t)\epsilon(t')} = \int d\omega\ S(\omega) e^{i
  \omega(t-t')} $%
. We assume a high frequency cutoff of the noise at $\gamma \ll E_1 -
E_0, \omega_0$.  The contribution to error at first order arises from
$\omega_{10}(t) = \omega_{10} + \epsilon(t) (\partial \omega_{10}
/ \partial \Delta_0)$%
.  When the pre-factor $\eta_{\Delta} = |\partial \omega_{10}/\partial
\Delta_0|^2$ is non zero, there is first-order dephasing.  In the
rotating frame, the off-diagonal density matrix element evolves as
\begin{equation} \rho_{01}(t) \sim 
  \exp\left[- \eta_{\Delta_0}\int d\omega\ S(\omega) \frac{\sin^2
      (\omega \tau/2)}{(\omega/2)^2}\right]\ . \label{e:decay}
\end{equation}
For example, when $\gamma \ll 1/\tau$, the decay is Gaussian with a
time constant $T_2 \sim T_{2,{\rm bare}}/\sqrt{\eta_{\Delta_0}}$%
, where the characteristic time $T_{2,{\rm bare}} = 1/\sqrt{\int
  d\omega\ S(\omega)}$. Bare dephasing times of $\sim 1$ ns were
observed for qubit frequencies of $\sim 2 \pi \times 20$
GHz~\cite{hayashi03,petta04}.  However, longer $T_{2,{\rm bare}}$
times might be achieved for qubits at $\sim 2 \pi \times 1$ GHz
through better high- and low-frequency filtering of electronics noise.
We will take $T_{2,{\rm bare}} = 10$ ns for the remainder of the
paper; this limit arises from the combined effect of charge-based
terms and nuclear spin-related
dephasing~\cite{bracker05,koppens05,petta05}.

\begin{figure}
\includegraphics[width=3.0in]{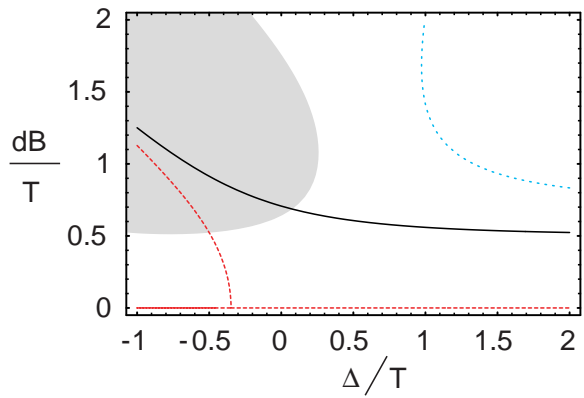}
\caption{
Solutions of $\partial
\omega_{10}/\partial \Delta_0=0$ (black), $\partial
\omega_{10}/\partial t=0$ (blue), and $\partial
\omega_{10}/\partial dB=0$ (red).  The region of $g_{\rm eff}/g > 0.3$
is marked in grey.
\label{f:dephasing}}
\end{figure}

At the zero derivative point ($\partial \omega/\partial \Delta_0 = 0$)
second-order terms must be considered~\cite{vion02,falci05,meier05}.
To the leading order~\cite{falci05}, we find dephasing occurs over a
timescale $T_2 \sim \omega_{10} (T_{2,{\rm bare}})^2$.  More detailed
calculations indicate that the underlying physics of the bath becomes
important near zero-derivative points~\cite{falci05,meier05}.  Thus
the optimal point combines the maximal coupling $g_{\rm eff}$ to the
resonator with an extended $T_2$ time (Fig.~\ref{f:dephasing}). In
this regime, with $\omega_{0} = 2 \pi \times 1.5$ GHz, $g_{\rm eff}
T_2 \approx 100$.  We note that away from the optimal point, even
assuming maximal coupling, $g_{\rm eff} T_2 \lesssim 1$.  For example,
in the strongly biased regime ($\Delta/t \gg 1$), $g_{\rm eff} T_2
\approx
g T_{2,{\rm bare}} \lesssim 1$.  We conclude that only near the
optimal point may the strong coupling regime be achieved. 

\section{Quantum control}
We now discuss potential applications.  Coherent control of the system
with little sensitivity to low frequency noise is achieved by driving
the resonator on resonance with a microwave pulse.  We can describe
this scenario by making the cavity state a coherent state,
$\ket{\alpha}$, letting us replace $\hat{a}$ with $\alpha$.  We would
expect Rabi oscillations between $\ket{0}$ and $\ket{1}$ with Rabi
frequency $\Omega = g_{\rm eff} \alpha$.  Low frequency components of
the driving field (DC offsets) are greatly reduced because of the gap
induced by $t$ and the insensitivity of the gap energy $\omega$ on
variations of $\Delta$.

Another technique is a quantum non-demolition measurement.  When the
microwave transition is detuned from the cavity by $\tilde{\Delta}$,
the evolution yeilds a qubit-dependent phase of $\phi_{QND} = \pm g^2
\tau /\tilde{\Delta}$ (for $\ket{0},\ket{1}$ states respectively) for
an injected microwave field over a time $\tau$ (limited by qubit
decay).  Assuming that the noise level of the detector exceeds the
background by about a factor of ten, the fidelity of such a QND
measurement of the encoded spin qubits is given by $F=1-10 \kappa /g^2
T_1$~\cite{blais04}.  This single qubit QND measurement can be
completed in a time $1/\kappa \sim 1\ \mu$s.

In addition, a wide variety of cavity QED quantum control techniques
may be accessible.  As an example, we focus on the SWAP operation, where the
qubit state is ``swapped'' with a photonic state of the resonator.
When there are no photons in the cavity and $\delta=0$, the state
$\ket{0} \ket{0}_{\rm cav}$ is stationary, while $\ket{1}\ket{0}_{\rm
  cav}$ oscillates with Rabi frequency $g_{\rm eff}/\hbar$ to the
state $\ket{0} \ket{1}_{\rm cav})$ (we use $\ket{n}_{\rm cav}$ to
indicate $n$ photons in the cavity mode).  When the time spent
oscillating is $\pi \hbar/g_{\rm eff}$, a quantum state of the
singlet--triplet system is mapped to the existence or absence of a
cavity photon:
\begin{equation}
  (\alpha \ket{0} + \beta \ket{1}) \ket{0}_{\rm cav} \rightarrow 
  \ket{0} (\alpha \ket{0} + i \beta \ket{1})_{\rm cav}\ .
\end{equation}
This process can be controlled by rapidly changing $\delta$ to and
from zero.  In essence, we can convert one quantum bit (the double-dot
system) to another quantum bit (the cavity photon), which can then be
used for a variety of quantum information tasks, such as long-distance
quantum gates, quantum communication, and quantum
repeaters~\cite{cirac95,childs00,schmidtkaler03}.  Furthermore, it may
allow for coupling to other qubit systems, such as atoms,
molecules~\cite{sorensen04}, or superconducting qubits~\cite{blais04}.

As an example, we now detail the expected errors for the SWAP
operation.  Both $T_1$ and $\kappa$ contribute to the decay of the
system, which lead to population of the state $\ket{0}\ket{0}$, while
dephasing, $T_{2}$, leads to phase errors in the transformation.
Including only relaxation terms, yields a probability of error in
``SWAP'' of $p_{T_1} = \pi \frac{\kappa + 1/T_1}{g_{\rm eff}}$.  In
addition, pure dephasing ($T_{2}$) leads to an error with probability
$p_{T_{2}} = \pi^2 / (T_{2} g_{\rm eff})^2$.  The optimal point has
$g_{\rm eff} \approx 0.4 \eta E_{10}$, $1/T_1 \approx 0.2
\Gamma(E_{10})$, and $T_{2,a} \approx \omega_{10} T_{2,{\rm bare}}^2$
from noise in $\Delta_0$ and $dB$.  Optimizing against both noise
sources suggests that for a bare dephasing time of $T_{2,{\rm bare}} =
10$ ns, a cavity frequency of $\sim$ 1.5 GHz is optimal for SWAP
operations, with a probability of error $\approx 2 \%$.  

\section{Conclusion}
In summary, we have shown how spin states of double quantum dots can
be coupled directly to microwave resonators.  An optimal operating
point with maximum coupling to the resonator and minimum coupling to
some sources of charge and spin noise is found.  This suggests a
variety of powerful quantum control techniques may become possible for
such a system.  

We thank A. Imamoglu, G. Burkard, and L. Childress for helpful
discussions.  This work was supported by ARO, NSF, the Sloan
Foundation, and the David and Lucille Packard Foundation.


\end{document}